\documentclass[prb,superscriptaddress, twocolumn]{revtex4-2} 
\usepackage{amsmath}
\usepackage{amsfonts}
\usepackage{graphicx}% Include figure files
\usepackage{graphicx, nicefrac}
\usepackage{dcolumn}% Align table columns on decimal point
\usepackage{float}
\usepackage{subfig}
\usepackage{bm}% bold math
\usepackage{braket} %para poner bras y kets
\usepackage{svg}
\usepackage{float}

\begin{document}

\title{Polarization vs. magnetic field: competing eigenbases in laser-driven atoms}
\author{Nicol\'as A. Nu\~nez Barreto}
\affiliation{Universidad de Buenos Aires, Facultad de Ciencias Exactas y Naturales, Departamento de Física,
Laboratorio de Iones y Átomos Fríos, Pabellón 1, Ciudad Universitaria, 1428 Buenos Aires, Argentina}
\affiliation{CONICET - Universidad de Buenos Aires, Instituto de Física de Buenos Aires (IFIBA), Pabellón 1,
Ciudad Universitaria, 1428 Buenos Aires, Argentina}
\author{Cecilia Cormick}
\affiliation{Instituto de F\'isica Enrique Gaviola, CONICET and Universidad Nacional de C\'ordoba,
Ciudad Universitaria, X5016LAE, C\'ordoba, Argentina}
\author{Christian T. Schmiegelow}
\affiliation{Universidad de Buenos Aires, Facultad de Ciencias Exactas y Naturales, Departamento de Física,
Laboratorio de Iones y Átomos Fríos, Pabellón 1, Ciudad Universitaria, 1428 Buenos Aires, Argentina}
\affiliation{CONICET - Universidad de Buenos Aires, Instituto de Física de Buenos Aires (IFIBA), Pabellón 1,
Ciudad Universitaria, 1428 Buenos Aires, Argentina}

\date{\today}

\begin{abstract}

We present experimental results and a theoretical model that illustrate how competing eigenbases can determine the dynamics of a fluorescing atom. In the absence of a magnetic field, the atom can get trapped in a dark state, which inhibits fluorescence. 
In general, this will happen when the magnetic degeneracy of the ground state is greater than the one of the excited state.  A canonical way to avoid optical pumping to dark states is to apply a magnetic field at an angle with respect to the polarization of the exciting light. 
This generates a competition of eigenbases which manifests as a crossover between two regimes dominated either by the laser or the magnetic field.
We illustrate this crossover with fluorescence measurements on a single laser-cooled calcium ion in a Paul trap and find that it occurs at a critical laser intensity that is proportional to the external magnetic field. 
We contrast our results with numerical simulations of the atomic levels involved and also present a simple theoretical model that provides excellent agreement with experimental results and facilitates the understanding of the dynamics.

\end{abstract}

\maketitle

\section{Introduction}

Experiments involving the manipulation of trapped atoms by means of lasers regularly make use of magnetic fields to set a ``quantization axis'' for the sublevels within electronic manifolds~\cite{chen2002quantization}. However, a magnetic field is neither necessary for the quantum treatment of the system nor for the labeling of the states. Instead, the introduction of a magnetic field can qualitatively change the evolution of the driven atom. In particular, the magnetic field, via the Zeeman effect, establishes energy differences between otherwise degenerate sublevels, with the field direction setting the quantization axis for the Zeeman eigenstates. Without these energy splittings, the atom can be optically pumped into so-called ``dark states''.  

\begin{figure}[ht]
\includegraphics[width=\columnwidth]{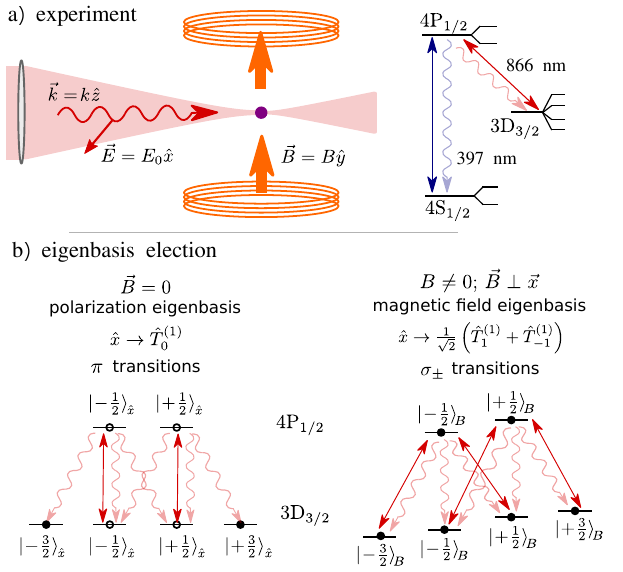}% Here is how to import EPS art
%\vspace*{-6mm}
\caption{\label{fig:experiment} a) Sketch of the experimental setup: transitions between electronic levels are driven by co propagating infrared and ultraviolet lasers, both linearly polarized in the same direction, perpendicular to the magnetic field $\vec B$. A scheme of the relevant levels in the calcium ion together with the laser transitions in our experiment is shown at the right. Within each manifold, the sublevels are split via the Zeeman effect. b) Electronic dynamics in the D-P transition illustrating the suppression of pumping to dark states. At the left, in the absence of magnetic field, the interaction with the laser takes the system to states which are not repumped into the fluorescence cycle. To the right, in presence of sufficiently large Zeeman splittings, the relevant basis is no longer the one set by the laser polarization, leading to continuous fluorescence.}
\end{figure}

In some situations, the appearance of dark states is a desirable feature, as it can be used to optically pump the electron into a particular sublevel~\cite{kastler1957optical,cohen1966optical,happer1972optical}. However, if one requires a steady fluorescence, as is the case when performing fluorescence measurements, pumping into dark states must be avoided. A canonical way to deal with this issue is to apply a magnetic field at an appropriate angle with respect to the polarization of the fluorescence field, generating mixing of bright and dark states and assuring a steady state fluorescence. We note that there are other alternative methods to avoid dark states, which include using unpolarized light or switching polarization dynamically in time~\cite{berkeland2002destabilization,lindvall2012dark,lindvall2013unpolarized,fordell2015broadband,xia2021destabilization}. Nevertheless, they tend to be experimentally more challenging and are beyond the scope of our discussion. 

Our experiment explores the suppression of dark states by means of a magnetic field monitoring the fluorescence of a single trapped ion as illustrated in Fig.~\ref{fig:experiment}~a). The underlying mechanism is a mixing of levels due to a choice of magnetic field with an eigenbasis which is different from the laser basis, as we describe in detail in this article. 
When one adds the magnetic field to the light-atom interaction, one finds a competition between two different eigenbases depending on the relative magnitudes of the magnetic field and the laser intensity. For vanishing magnetic field, the system eigenbasis is the one set by the linear polarization of the laser. In this ``polarization eigenbasis'' the lower manifold of our ion always contains dark states, as illustrated in the left panel of Fig.~\ref{fig:experiment}~b). In the presence of a magnetic field $\vec B$, the polarization-eigenstates will no longer be energy eigenstates, since they will not coincide with the Zeeman levels determined by the magnetic field direction. Then, the polarization-eigenstates will mix at the Larmor frequency. 

Alternatively, one can see why dark states are eliminated by the inclusion of the magnetic field by considering the transitions in the basis of the Zeeman eigenstates, as illustrated in the right panel of Fig.~\ref{fig:experiment}~b). In this basis, the transitions generated by the laser are not $\pi$, but $\sigma^+ + \sigma^-$~transitions, which connect each of the ground states with an excited one, generating steady-state fluorescence. This happens when the leading eigenbasis is the one determined by the magnetic field, i.e. when the Larmor frequency is sufficiently large for a given laser intensity, or equivalently when the laser pump is weak enough for a given magnetic field. 

The above discussion suggests one can evidence the competition between the two different eigenbases by monitoring the intensity of the light emitted by the atom as a function of the intensity of the laser field for a fixed magnetic field. At low laser powers, when the Zeeman splitting dominates, the relation between laser intensity and fluorescence is linear. As the laser power is increased, one usually expects the emission to saturate approaching a maximum value determined by the spontaneous emission rate. Instead, one observes that, before reaching saturation, the fluorescence peaks and starts to decrease, getting close to zero when approximately dark states appear. The value of laser intensity at the fluorescence peak increases as the magnetic field strength is increased.

In the following, we study both theoretically and experimentally this transition between the two regimes, dominated either by the magnetic field $\vec{B}$ or by the laser field $\vec{E}_{laser}$.
We present an experimental investigation of this phenomenon for a dipolar transition of trapped calcium ions, confirming the presence of a fluorescence maximum as the laser intensity is varied. We also observe that the location of this maximum depends on the strength of the magnetic field, in agreement with the description above. We then provide a simplified theoretical treatment which reproduces the main features of the system and leads to a better understanding of the dynamics. 

The paper is organized as follows: 
Sec. \ref{sec:dark} provides a short introduction to optical pumping into dark states in connection with the dimension of the atomic levels. In Sec. \ref{sec:setup} we briefly describe our experimental setup. 
In Sec. \ref{sec:results} we present the results of the experiment and discuss the physics involved. In Sec. \ref{sec:model} we provide analytical calculations for a simple model and compare them with the experimental results. Finally, in Sec. \ref{sec:conclusions} we summarize our work. An Appendix is included providing further experimental details.

\section{Dark states and electronic degeneracies}
\label{sec:dark}

\begin{figure}[t]
\includegraphics[width=0.95\columnwidth]{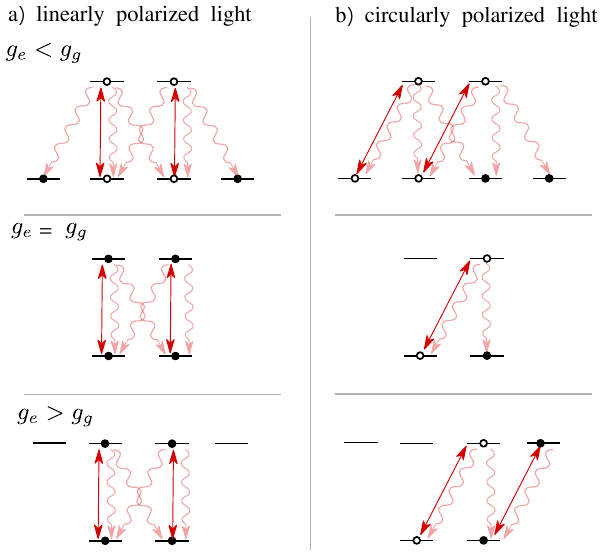}% Here is how to import EPS art
\caption{\label{fig:pumping-schematic} Schematic representation of electronic levels illustrating optical pumping to dark states. (a) and (b) show a dipolar transition driven by $\pi$ and circularly polarized light, respectively. Spontaneous decay is indicated with curvy lines. If the excited state has a lower degeneracy than the ground state, as shown in the upper row, the system is always pumped into dark states. If the degeneracy of the ground state is lower, corresponding to the bottom row, there are no dark states and the system continuously fluoresces. The intermediate case, when both excited and ground manifolds have the same degeneracy, can exhibit dark states or not depending on the laser polarization.}
\end{figure}

The appearance of dark states 
in Zeeman-degenerate sublevels 
is generic in atomic systems which have higher degeneracies in the lower manifold than in the excited one. We illustrate this idea in Fig.~\ref{fig:pumping-schematic}~a) where we consider transitions between states with half-integer total angular momentum (a similar analysis can be carried out for integer angular momentum). The upper row shows a case where the lower state has larger degeneracy ($g_g>g_e$), namely we consider that the total angular momentum of the ground state is $J_g=3/2$ and that of the excited state is $J_e=1/2$. In the absence of Zeeman shifts, for any fixed laser polarization one can find a basis such that some ground sublevels are decoupled from the laser interaction. When the atom is excited, it can decay by spontaneous emission into any of the ground states, including those not coupled to the laser, which will not be cycled back to the excited state. When this happens, we say the atom is pumped into a dark state, and fluorescence stops. 

In the upper level of Fig.~\ref{fig:pumping-schematic} the driving field can be assumed to be $\pi$-polarized, but any other laser polarization leads to the same scenario. For linear polarization, the natural axis for the labeling of the states is along the polarization direction of the laser $\hat\epsilon$. For circular $\sigma^{\pm}$ polarizations, the direction chosen for the labeling of the states is given by the propagation direction of the laser $\vec k$. Any other polarization, as long as it is a pure polarization, will generate the same behavior when choosing the basis appropriately: dark states will always appear when the degeneracy of the excited level is smaller than the one of the ground manifold. 

On the contrary, no dark states appear when the upper level has a higher degeneracy than the lower one ($g_e>g_g$). This is shown in the lower panel of Fig.~\ref{fig:pumping-schematic}, inverting the degeneracies of the upper and lower levels with respect to the previous case. Now, spontaneous decay always bring the electron back to a level which will continue to cycle to the excited state. This will occur in general for any system where the degeneracy of the ground state is smaller than the one of the upper level. Even for circularly polarized light, it is straightforward to check that the electron gets optically pumped to an extreme magnetic state, but will still fluoresce. We stress that the concept of optical pumping in this context refers to the fact that the population of the atomic levels is strongly affected by the laser field, but does not necessary imply the existence of dark states.

The intermediate case, where the degeneracies of the ground and excited states are equal ($g_e=g_g$), can either exhibit dark states or not depending on the polarization of the exciting beam. For instance, as seen in the middle panel of Fig.~\ref{fig:pumping-schematic}, for $\pi$-polarization there is a continuous fluorescence cycle, while  circular polarization  generates optical pumping to dark states. 

The main focus of our work is the analysis of the competition between laser driving and magnetic field underlying the appearance or suppression of dark states. Thus, we consider only the case depicted in the upper panel of Fig.~\ref{fig:pumping-schematic}, corresponding in our experiment to the 3D$_{3/2}$ and 4P$_{1/2}$ manifolds of the calcium ion as sketched in Fig.~\ref{fig:experiment}~a). Further experimental details are provided in the next Section.

\section{Experimental Setup} \label{sec:setup}

To illustrate the competition between the two eigenbases associated with laser polarization and magnetic field we examine the fluorescence of a single trapped calcium ion as a function of the intensities of the driving laser and the magnetic field. The relevant levels of $^{40}$Ca$^+$ are shown in Fig~\ref{fig:experiment}~a): the doubly degenerate ground state $4$S$_{1/2}$ is connected via a dipole transition near 397~nm to the excited $4$P$_{1/2}$ state, which is also doubly degenerate. This excited state is dipole-connected to a lower-lying metastable $3$D$_{3/2}$ state, which has four-fold degeneracy. Depending on the driving laser fields, one can find dark states in the S or D manifolds, in both at the same time, or in none of them. Here, we concentrate on the appearance of dark states in the D level, as a function of the intensity of the infrared (IR) field driving the D-P transition near 866~nm. The transition between S and P is simultaneously driven with an ultraviolet (UV) linearly polarized laser to repump the IR cycle of interest, preventing the atom from falling into dark states of the S manifold. 

From now on we focus on the dynamics of the IR transition. The polarization is set linear, and at 90$^\circ$ with respect to the magnetic field. When the magnetic field basis dominates, the laser polarization is seen by the atom as a combination of $\sigma^+$ and $\sigma^-$ polarization, driving $\Delta m=\pm 1$ transitions, as illustrated in the right panel of Fig.~\ref{fig:experiment}~b). This allows depopulation of all states in the D manifold to keep the ion fluorescing. In the opposite limit, when the magnetic field is very weak with respect to the laser intensity, the preferred eigen-direction is the one set by the linear polarization. Then the field is seen by the atom as $\pi$ polarization, which only drives $\Delta m=0$ transitions. As shown in the left panel of Fig.~\ref{fig:experiment}~b), the electron then eventually decays into one of the $m=\pm$3/2 states, and fluorescence is suppressed as described above.

\begin{figure*}[th]
%\includegraphics[]{Figuras/theshold.png}
%\centering
\includegraphics[width=0.9\textwidth]{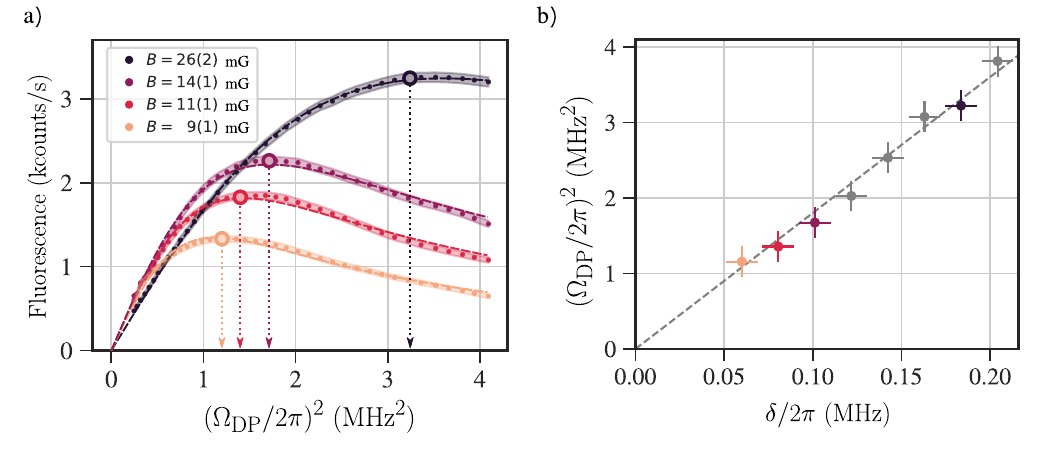}
\caption{\label{fig:02_measurements} a) Measurement of the fluorescence counts versus squared Rabi frequency of the D-P transition, proportional to the optical power of the IR laser, for four different values of the magnetic field (stronger fields corresponding to darker curves). The experimental data (dots) are plotted
with their uncertainty (shaded area). The detuning $\Delta$ of the IR laser is fixed to close to zero. In each curve, a maximum fluorescence value can be appreciated, such that after this point the fluorescence decreases with increasing optical power. We also plot a dashed line over each curve, which represents a fit to the 4-level simplified model of Sec.~\ref{sec:model}. b) Plot of the squared Rabi frequency at maximum fluorescence, proportional to the intensity of the laser, versus the Larmor frequency $\delta/2\pi$, proportional to the strength of the magnetic field. A linear behaviour between the two magnitudes can be appreciated in this regime. Along with the measurements we plot with a dashed grey line the theoretical curve using a full 8-level system treatment for $\Delta=0$, as explained in the appendix, which is in good agreement with the measurements. }
\end{figure*}

We monitor the transition between the two dominating eigenbases by recording the emitted fluorescence as a function of the power of the IR laser at the relevant D-P transition for various choices of magnetic field intensity. In all experiments we observe fluorescence on the S-P transition near 397~nm, which only occurs if the population is not pumped into dark states. The UV laser is tuned half linewidth to the red and set to have linear polarization: this keeps the ion cold below 5~mK via Doppler cooling while avoiding optical pumping onto dark S states. Fluorescence is collected with a 50~mm objective and recorded using a photomultiplier tube. Further experimental details which are not essential for the understanding of the main body of this work are provided in Appendix~\ref{sec:appendix} and in Ref.~\cite{barreto2022three}.

\section{Results} \label{sec:results}

The main results of our experimental exploration are shown in Fig.~\ref{fig:02_measurements}. Panel a) shows the total fluorescence collected as a function of the squared of the Rabi frequency, proportional to the laser intensity, for different values of the magnetic field. We also plot for each curve a fit to a simplified theoretical model using a 4-level system, explained in the next section, finding very good agreement for the functional dependence of these curves. We note that to calibrate the horizontal axis as a Rabi frequency, and to obtain the magnetic field of each curve, we fit atomic spectra as a function of the IR detuning $\Delta$ with an 8-level model as explained in the Appendix. 

We see in the plots that for low powers of the IR laser, the fluorescence grows linearly with the laser power. This is the expected behavior for a system well below the saturation intensity and where there are no dark states. Here, the magnetic field basis dominates and there is no optical pumping.  Above some power threshold, signalled with a circle and a dashed line for each curve, we see that the fluorescence starts to decay with increasing laser intensity. We find that this point depends on the value of the external magnetic field: the higher the magnetic field, the  higher the laser intensity at the turning point. 
The maxima of each curve allows one to identify a threshold point between the Zeeman eigenbasis (weak laser power) and the laser eigenbasis (strong laser power). 

As a next step, we study the dependence of this threshold power as a function of the magnetic field magnitude $B$. The latter is expressed in terms of Larmor frequency of the D states defined as $\delta/2\pi = g_L\,(\mu_B/h)B$, where $\mu_B$ is the Bohr magneton, $h$ is the Planck constant and $g_L$ is the Land\'e factor of the D states, which in this case is equal to $4/5$. This magnitude corresponds to the Zeeman splitting of the D states. In Fig.~\ref{fig:02_measurements}~b) we show the results, which indicate a linear dependence of the threshold power with the magnetic field. Here, the dashed line are the results obtained from a numerical simulation, where we resort to the optical Bloch equations of the full 8-level system as treated in~\cite{barreto2022three}. As seen, the full simulation shows excellent agreement with the measurements.  The simplified theoretical model of the next Section, reproduces the observed results qualitatively very well too, predicting this same linear behaviour, but with a different slope. 

From these results we confirm that one can identify a transition between the regimes where either the laser-defined basis or the magnetic field basis play a dominant role in the dynamics. We also observe that there is a linear relation between the magnitude of the magnetic field and the laser power at which one finds the turning point in fluorescence. The reason for this linear behaviour is less obvious: a competition between characteristic frequencies could make one expect a linear relation between the Zeeman splitting (Larmor frequency) and the laser amplitude ($\Omega_{\mathrm{DP}}$) instead of the laser intensity ($\Omega_{\mathrm{DP}}^2$). Understanding this point requires a more mathematical approach. In the next Section we provide a simplified model that captures the essential features observed.

\section{Simplified theoretical model} \label{sec:model}

\begin{figure}[h]
\includegraphics[width=0.65\columnwidth]{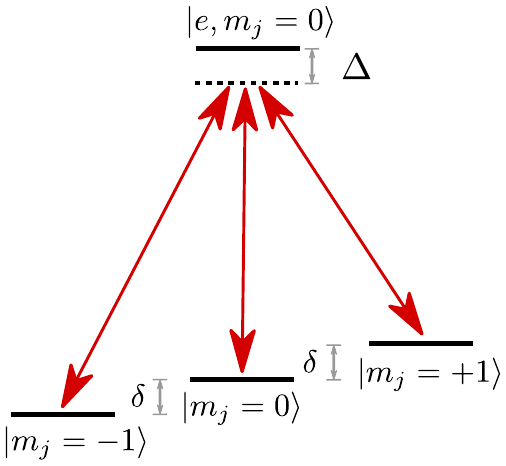}% Here is how to import EPS art
%\vspace*{-6mm}
\caption{\label{fig:05_4levels} Simplified model with only 4 levels to facilitate the theoretical analysis. The Zeeman energy splittings $\delta$ within the ground manifold are proportional to the magnetic field applied.}
\end{figure}

In this Section we describe a simplified model which can be solved analytically, allowing one to better understand the behaviour observed experimentally and numerically. To avoid exceedingly cumbersome algebra, we consider a 4-level system, with one excited state and three ground sublevels, as in Fig.~\ref{fig:05_4levels}. The ground sublevels are identified by the quantum numbers $m_j = \{ -1, 0, 1 \}$ in the basis determined by the magnetic field, whereas the excited state has $m_j=0$. For simplicity, we assume that all three transitions are driven with the same Rabi frequency $\Omega$. We also introduce a detuning $\Delta$ of the laser with respect to the electronic transition. The levels included are chosen to illustrate the dynamics involving the D$_{3/2}$ and P$_{1/2}$ manifolds only, with a reduced number of sublevels to facilitate the derivation of analytical results. % The absence of additional levels representing the S levels implies that our model will not exhibit dark resonances as in Fig.~\ref{fig:01_levels_cpttheo}, which are nevertheless not essential for our study.

In the frame rotating at the laser frequency, and choosing the excited state to be the last one in the basis, the Hamiltonian part of the evolution is given by \cite{steck2007quantum, lukinnotes}:
\begin{equation} \label{eq:Hamiltonian}
H=\hbar
\begin{pmatrix}
\Delta - \delta\,\, & 0 & 0 & \Omega/2\\
0 & \,\,\Delta\,\, & 0 & \Omega/2\\
0 & 0 & \,\,\Delta+\delta\,\, & \Omega/2\\
\,\Omega/2 & \,\Omega/2 & \,\Omega/2 & \,0
\end{pmatrix}.
\end{equation}
Here, $\delta$ corresponds to the Zeeman splitting of the ground state sublevels, as can be seen in Fig.~\ref{fig:05_4levels}, since we consider a simplified case where $g_L=1$. The full dynamics, including spontaneous emission, can be described by a Lindblad equation of the form:
\begin{equation}
    \frac{d\rho}{dt}=-\frac{i}{\hbar}\left[ H, \rho \right] + \mathcal{L}_\Gamma\rho
    \label{eq:master_eq}
\end{equation}
with $\mathcal{L}_\Gamma$ accounting for the decay from the excited to the ground levels. In turn, this contribution can be written as a sum of different phenomena:
\begin{multline}
   \mathcal{L}_\Gamma \rho =  - \frac{\Gamma_T}{2}\{|e\rangle\langle e|,\rho\} \, + \sum_{m=0,\pm1} \Gamma_m\, |m\rangle\langle e|\rho|e\rangle\langle m| \\
   + \Gamma_S |e\rangle\langle e| \rho |e\rangle\langle e| \,.
\end{multline}
Here, the curly brackets denote an anticommutator, and to simplify the notation, we use $e$ to label the excited state and $m=0,\pm1$ to label the ground sublevels. In this equation, $\Gamma_m$ represents the decay rate into each of the three sublevels in the D manifold, $\Gamma_S$ induces an effective dephasing through decay into the S manifold followed by repumping to the excited state, $\Gamma_T=\Gamma_S+\sum_m \Gamma_m$ is the total dephasing rate, and we neglect dephasing due to laser fluctuations.

In the expressions above, $\Omega$ is associated with the laser amplitude, so that for $\Omega\to0$ the fluorescence is negligible and the atom is always in the ground manifold. The magnitude of the magnetic field $B$ is quantified by the Zeeman splitting $\delta$ between ground sublevels. If $\delta=0$, the ground manifold is degenerate. Furthermore, because we chose for simplicity the case where the Rabi frequency is the same for all ground sublevels, for $\delta=0$ the Hamiltonian in Eq.~\eqref{eq:Hamiltonian} is invariant under permutations of these sublevels. Therefore, the laser only couples the excited state with the ground sublevel of the form $(\ket{0}+\ket{1}+\ket{-1})/\sqrt{3}$. By acting with the Hamiltonian on this state, one can easily check that the Rabi frequency for this transition is equal to $3\Omega$. This particular ground state which is driven by the laser is the ``bright state''. The ground states orthogonal to this one do not couple to the excited state via the laser drive, so we label them as dark states. Spontaneous emission, governed by the Lindblad term, can populate dark states. If this happens, the atom will remain in these states, since in absence of magnetic field they are eigenstates of the Hamiltonian. In this way, after a transient, only dark states will be populated. Thus, in both the limits of very weak laser and of very weak magnetic field the fluorescence will be negligible. We note that choosing different Rabi frequencies changes the form of the bright state, but not the qualitative behavior.

We now explore the intermediate regime when both $\delta$ and $\Omega$ are non-negligible. For this, one can solve for the asymptotic state of the equation of motion, Eq.~\eqref{eq:master_eq}, including coherent and dissipative dynamics, setting the time derivative of $\rho$ equal to zero. This leads to an inhomogeneous linear system of equations which can be solved in a straightforward manner. In particular, from the solution one can extract the asymptotic population of the excited state $p_e$, which is proportional to the total fluorescence at steady state.  To simplify things further, we consider the case when $\Gamma_m=\Gamma_D/3$ for all sublevels, with $\Gamma_D$ the total decay rate into the D manifold. This assumption simplifies the calculation and reproduces the general behavior, even if it does not represent the true decay rates. In this case, we obtain:
\begin{equation}
    p_e = \left\{ \frac{\Gamma_D}{\Gamma_T} \left[\frac{\Gamma_T^2+4\Delta^2+8\delta^2/3 }{\Omega^2} + \frac{ 3 \Omega^2 }{ 2\delta^2 } \right] + \frac{4\Gamma_S}{\Gamma_T} \right\}^{-1} \,.
    \label{eq:pe}
\end{equation}

Several conclusions can be drawn from this formula. 
First, we see that the asymptotic population of the excited state tends to zero for vanishing magnetic field ($\delta\rightarrow0$) or vanishing laser power ($\Omega\rightarrow0$). More precisely, we can see that the fluorescence signal is weak when  $2\delta^2\ll3\Omega^2$ (low magnetic field) or $\Omega^2\ll\Gamma_T^2+4\Delta^2+8\delta^2/3$ (low laser power). One can also analytically calculate the point of maximum fluorescence for varying $\Omega$, which gives
\begin{equation}
 \Omega^2_{\rm max} = \sqrt{\frac{2}{3}} \,|\delta|\, \sqrt{\Gamma_T^2+4\Delta^2+8\delta^2/3}\,.
 \label{eq:Omega_max}
\end{equation}
The case of interest to describe our experiment is the one of low $B$ field, $|\delta|\ll \Gamma_T$, implying that the transition linewidth is much broader than the Zeeman splitting. This means that the above formula can be approximated by
\begin{equation}
\Omega^2_{\rm max} = \sqrt{\frac{2}{3}} \,|\delta|\, \sqrt{\Gamma_T^2+4\Delta^2}\,,
\end{equation}
consistent with the linear relation between laser power and magnetic field observed in Fig.~\ref{fig:02_measurements}. We also notice that the slope of the linear relation depends on the value of the detuning $\Delta$.

We compare the measurements of Fig.~\ref{fig:02_measurements}~a) with the analytic expression of Eq.~\eqref{eq:pe}, adjusting by an overall scaling factor relating total and recorded fluorescence. We find very good qualitative agreement, as can be seen from the the dashed lines (model) which are shown over the data. We also compare the experimental value of the slope of Fig.~\ref{fig:02_measurements}~b), which is $17.9(2)~$MHz, with the value given by the simplified model of Eq.~\eqref{eq:Omega_max} for $\Delta=0$. The predicted slope, using the values of the linewidths from~\cite{hettrich2015measurement}, becomes $\sqrt{{2}/{3}}\,\Gamma_T / (2\pi)\sim 18.9~\mathrm{MHz}$. 
Both values have a very good level of agreement given the simplification of the model considered. They can be related by a factor of the order of 1 that accounts for the difference between the model and the real system, which involves for example Clebsch-Gordan coefficients and the Land\'e factors of the states.
For a more precise prediction we ran a a full simulation of the 8-level system which, as stated above, provides a value for the slope which is correct, within experimental error, as shown in Fig.~\ref{fig:02_measurements}~b).

%We attribute this slight difference to the simplification of the level system, which manages to capture the general behavior and the order of magnitude of the parameters but not their precise value. For a precise prediction we ran a a full simulation of the 8-level system which, as stated above, provides a value for the slope which is correct, within experimental error, as shown in Fig.~\ref{fig:02_measurements}~b).

Another qualitatively correct prediction of Eq.~\eqref{eq:pe} is given by the limit of large $|\Delta|$. For large detuning, the formula indicates a decay of the fluorescence proportional to $|\Delta|^{-2}$. Furthermore, it also predicts that the scale for this decay depends strongly on the values of the remaining parameters. Indeed, for low magnetic fields, ``large detuning'' actually means $4\Delta^2\gg \Gamma_T^2+3\Omega^4/\delta^2$. This $\delta^2$ in the denominator of the last term leads for low magnetic fields to a very slow decay of the fluorescence with increasing $|\Delta|$, in agreement with the spectra shown in the Appendix.

\section{Conclusions} \label{sec:conclusions}

We have experimentally characterized a transition in the behavior of the fluorescence spectrum corresponding to the onset of the ``laser-defined basis'', when the Zeeman splittings are small compared to the laser power. We have provided a conceptual description of this transition, supported by numerical simulations. The qualitative features of the phenomena studied are also captured by a reduced 4-level model which we solved analytically. We expect that this work can illustrate aspects of fluorescence spectroscopy which are rarely discussed in depth, but which are essential for a good understanding of atom-field interactions.

\medskip

\section{Acknowledgements}

C.C acknowledges funding from grant PICT 2020-SERIEA-00959 from ANPCyT (Argentina).
C.T.S. and N.N.B. acknowledge support for grants PICT2018-03350, PICT2019-04349 and PICT2021-I-A-01288 from ANPCyT (Argentina) and grant UBACYT 2018 Mod I - 20020170100616BA from Universidad de Buenos Aires, as well as generous support from F. Schmidt-Kaler.

\appendix

\section{Calibration of the experiment} \label{sec:appendix}

\begin{figure}[h]
%\includegraphics[]{Figuras/CPT_exp.pdf}
%\centering
{\includegraphics[scale=1]{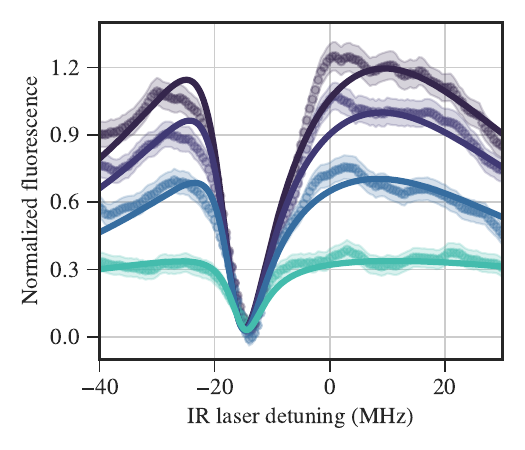} }%
\caption{\label{fig:04_bcalibration} Atomic spectra for magnetic fields of $\{ 39(1), 33(1), 23(1), 20(2) \}$~mG, with darker colors corresponding to stronger values of $B$. These estimates were obtained by sweeping the IR laser detuning around resonance while keeping the UV laser detuning fixed. The continuous lines are fits to the 8-level system. In all the cases the magnetic field is low enough that dark resonances between individual D and S sublevels cannot be resolved. Yet, from the fits the value of $B$ can be properly retrieved. Here, $\left( \Omega_{DP}/\Gamma_{DP}\right)^2=2.27\pm0.14$, which corresponds to an optical power of $\sim60$~$\mu$W. The UV laser has  $\left( \Omega_{SP}/\Gamma_{SP}\right)^2=1.19\pm0.04$, which corresponds to an optical power of $\sim15$~$\mu$W, and detuning $\Delta_{\mathrm{UV}}=(-14.7\pm0.2)~$MHz.}
\end{figure}

In order to calibrate experimental parameters such as magnetic field and Rabi frequencies, we resort to atomic spectra of the D-P transition. We keep fixed the frequency of the UV laser and vary the one of the IR field obtaining spectra like the ones shown in Fig.~\ref{fig:04_bcalibration} for four different magnetic fields. The deep in the left part is a dark resonance that arises due to coherent population trapping which involves mixtures of S and D sublevels~\cite{barreto2022three}. For strong magnetic fields, there is more than one deep depending on the polarization of the lasers. For each spectrum, the field $\vec B$ and the laser polarizations are kept constant, while the frequency of the IR laser is varied. The lowest value of $B$ studied was 9(1)~mG, since for lower fields the cooling is inefficient due to the low fluorescence rate.

\bibliography{bibpaper}

\end{document}